\documentclass[aps,prb,10pt,twocolumn,superscriptaddress,a4paper,longbibliography]{revtex4-2}
\usepackage[dvipdfmx]{graphicx}

\usepackage{amsmath,amsthm,amssymb}
\usepackage{bm}
\usepackage{color}
\usepackage{ifthen}
\usepackage[svgnames]{xcolor}
\usepackage{multirow,bigdelim}
\usepackage{comment}
\usepackage{url}
\usepackage{csquotes}
\usepackage{soul}

\usepackage[
pagebackref=false,
colorlinks=true,
linkcolor=blue,
urlcolor=blue,
filecolor=black,
citecolor=blue,
pdfstartview=FitV,
pdftitle={},
pdfauthor={},
pdfsubject={},
pdfkeywords={},
bookmarksopen=true
]{hyperref}
\usepackage{orcidlink}

\newcommand{\1}{\mbox{1}\hspace{-0.25em}\mbox{l}}

\def\fivequad{\quad\quad\quad\quad\quad}

\def\det{\mathrm{det}}

\newcommand{\pmat}[1]{\begin{pmatrix} #1 \end{pmatrix}}

\newcommand{\ii}{\mathrm{i}}

\newcommand{\sectitle}[1]{\emph{{#1}.---}}

\begin{document}

\title{
Nonlinear Frequency-Momentum Topology and Doubling of Multifold Exceptional Points
}
\author{
Tsuneya Yoshida
\orcidlink{0000-0002-2276-1009}
}
\email{yoshida.tsuneya.2z@kyoto-u.ac.jp}
\affiliation{%
Department of Physics, Kyoto University, Kyoto 606-8502, Japan
}%

\date{\today}

\begin{abstract}
Even in the linear limit, the topology of multifold (also called higher-order) exceptional points across the Brillouin zone has lacked a general characterization, leaving the doubling theorem essentially limited to two-fold exceptional points.
Here, we establish the doubling theorem of $n$-fold exceptional points [EP$n$s ($n=2,3,\ldots$)] for systems where nonlinearity enters through eigenvalues. To this end, we introduce new topological invariants, termed frequency-momentum winding numbers, which characterize nonlinear EP$n$s in $m$-band systems throughout the Brillouin zone for arbitrary $n$ and $m$ ($m\geq n$). 
These invariants enable a unified proof of the doubling theorem in the absence of symmetry and under several symmetry constraints, including parity-time ($PT$) and charge-conjugation-parity symmetries. 
Furthermore, even in the linear limit, the frequency-momentum winding number indicates $\mathbb{Z}$ topology of $PT$-symmetric EP$2$s which is beyond the previously reported $\mathbb{Z}_2$ topology. The frequency-momentum winding numbers can also be extended to a class of coupled resonators in which nonlinearity enters via the eigenvectors, whereas the spectrum is determined by a nonlinear scalar equation for the frequency.
\end{abstract}

\maketitle

\sectitle{Introduction}
%
Doubling of topological quasiparticles, enforced by the vanishing total topological charge in the Brillouin zone (BZ), plays a fundamental role in topological physics~\cite{Nielsen_PhysLettB1981,Nielsen_NuclPhysB1981_I}. A representative example is provided by Weyl semimetals~\cite{Murakami_NJP2007,Wan_PRB2011,Burkov_PRL2011,Bevan_Nature1997,Armitage_RMP2018}, which host pairs of Weyl nodes, i.e., linear band-touching points with quantized Chern number. The doubling of Weyl nodes underlies characteristic phenomena such as surface Fermi arcs~\cite{Xu_Science2015,Huang_NatComm2015,Weng_PRX2015,Silaev_PRB2012} and anomalous transport responses~\cite{Nielsen_PhysLettB1983,Yang_PRB2011,Burkov_PRL2014,Liu_NatPhys2018,Nakatsuji_Nature2015,Wang_NatComm2018,Son_PRB2013,Huang_PRX15,Zhang_NatComm2016,Xiong_Science2015}. 
Recently, extensions of the doubling theorem to exceptional points in non-Hermitian systems have been explored~\cite{Yang_PRL2021,Yoshida_PRR2025}. At exceptional points, the coalescence of eigenvalues and eigenvectors leads to a dispersion relation with a fractional exponent~\cite{Berry_CzeJPhys2004,WDHeiss_JPhysAMathGen2004,Rotter_JPhysAMathTheo2009,Shen_PRL2018,Kawabata_PRL2019,Kozii_PRB2024,Bergholtz_RMP2021,Ashida02072020}, which is generalized to multifold (or higher-order) exceptional points~\cite{Heiss_JPhysAMathTheo_2012,Demange_JPhysAMathTheo2012,Delplace_PRL2021,Mandal_PRL2021,Sayyad_PRRes2022,Montag_PRRes2024,Yoshida_PRR2025}. The doubling of two-fold exceptional points is proved~\cite{Yang_PRL2021} and accounts for bulk Fermi arcs~\cite{Kozii_PRB2024,Doppler_Nature2016,Zhou_Science2018,Tang_PRL2021,Hofmann_PRR2020,Gao_Nature2015,Yoshida_PRB2018,Yoshida_PRB2019I,Zyuzin_PRB2018}.

While the above phenomena are described in terms of eigenvalue equations, topological physics has recently been extended to nonlinear systems, including classical metamaterials~\cite{Leykam_PRL2016,Benzaouia_APL2022,Veenstra_Nature2024,Smirnova_APR2020,Xie_MINE2026,Sahin_AdvSci2025,Kuzmiak_PRB1994,Chern_PRE2006,ZWang_PRL2008,Huang_IntJEngSci2009,SHLee_PRB2016,Ozawa_RMP2019}
and interacting quantum matters~\cite{Gulevich_SciRep2017,Mostaan_NatComm2022,Kozii_PRB2024,Yoshida_PRB2018}.
In such systems, nonlinearity can enter either through the eigenvectors~\cite{Sone_PRR2022,Sone_NatPhys2024,Sone_CommPhys2025,Sone_PRR2026,Jezequel_PRB2022,Schindler_PRB2025}
or through the eigenvalues~\cite{Isobe_PRL2024,MSilveirinha_PRB2015,Raman_PRL2010,Feng_SciRep2025,Xiao_PRL2016}. 
Nonlinearity of eigenvectors affects the stability of edge states, giving rise to nonlinear excitations such as chiral solitons~\cite{Ablowitz_PRA2014,Gulevich_SciRep2017,Snee_EML2019,Ezawa_PRB2022II}, while nonlinearity of eigenvalues effectively increases the number of bands~\cite{Raman_PRL2010,Feng_SciRep2025}.
Nonlinear topology has been further explored in non-Hermitian systems~\cite{Bai_PRL2023,Bai_NatlSciRev2022,Fang_PRB2025,Wingenbach_arXiv2025,Ezawa_PRB2022,Castro_PRB2025,Kawabata_PRL2025,Hamanaka_arXiv2026,Yoshida_PRB2025,Feng_SciRep2025}, particularly through the interplay between nonlinearity and exceptional points. 
The nonlinearity of eigenvalues can induce exceptional points~\cite{Yoshida_PRB2025} and even increase their multiplicity~\footnote{
For a Lorentz dispersive medium, Ref.~\cite{Feng_SciRep2025} maps a system with nonlinearity of eigenvalues to an effective linear system.
}~\cite{Feng_SciRep2025,Bai_PRL2023,Bai_NatlSciRev2022,Fang_PRB2025}.

Despite the above progress, the doubling theorem of exceptional points in nonlinear systems has yet to be established, especially for $n$-fold exceptional points (EP$n$s). 
The central obstacle is the absence of topological invariants that characterize EP$n$s in nonlinear systems with an arbitrary number of bands. 
Furthermore, even in the linear limit, no topological invariant is known to characterize EP$n$s ($n\geq 3$) across the BZ in $m$-band systems~\cite{Yoshida_SciPostPhys2026}.
As a consequence, the existing doubling theorem is restricted to EP$2$s in the linear regime~\cite{Yang_PRL2021}.

We hereby establish the doubling theorem of EP$n$s with arbitrary $n$ ($n\geq 2$) for systems where nonlinearity enters through eigenvalues (i.e., frequency). 
To this end, we introduce new topological invariants, termed frequency-momentum (FM) winding numbers, which characterize nonlinear EP$n$s in $m$-band systems throughout the BZ for arbitrary $n$ and $m$. 
The FM winding numbers enable a proof of the doubling theorem both in the absence and presence of symmetry, including parity-time ($PT$) and charge-conjugation-parity ($CP$) symmetries.
In particular, they imply the doubling of EP$3$s in the two-dimensional BZ and EP$4$s in the three-dimensional BZ under $PT$ symmetry, and of EP$5$s and EP$7$s when both $PT$ and $CP$ symmetries are imposed.
Furthermore, even in the linear limit, the FM winding numbers reveal previously unrecognized stability of EP$2$s with $PT$ symmetry; our invariant indicates $\mathbb{Z}$ topology refining the existing $\mathbb{Z}_2$ topological characterization.
The FM winding numbers can also be extended to a class of coupled resonators with nonlinearity of eigenvectors, whose spectrum is determined by a nonlinear scalar equation for frequency.

\sectitle{Frequency-momentum winding number}
%
We consider a system described by a nonlinear eigenvalue equation
\begin{align}
\label{eq: F psi=0}
F(\omega_n,\bm{k}) |\psi_n(\omega_n,\bm{k})\rangle &= 0,
\end{align}
where $F(\omega_n,\bm{k})$ is an $N \times N$ matrix, $|\psi(\omega_n,\bm{k})\rangle$ are eigenvectors, and $\omega_n \in \mathbb{C}$ are eigenvalues ($n=1,2,\ldots$). 
The real vector $\bm{k}$ denotes momentum (or wavenumber) in the $D$-dimensional BZ.
The eigenvalues $\omega_n$ are obtained by solving 
\begin{align}
\label{eq: det F=0}
\mathrm{det}F(\omega_n,\bm{k}) &= 0.
\end{align}
%
We suppose that $\det F(\omega,\bm{k})$ becomes independent of $\bm{k}$ for large $\omega$, as exemplified by 
$\det F(\omega,\bm{k})\sim \omega^r$ for $|\omega|\to \infty$ with integer $r$.
Equations~\eqref{eq: F psi=0} and \eqref{eq: det F=0} describe band structures of metamaterials with dispersive media~\cite{Kuzmiak_PRB1994,Chern_PRE2006,Huang_IntJEngSci2009,SHLee_PRB2016,Raman_PRL2010,Feng_SciRep2025}
and quasiparticles in quantum systems~\cite{Kozii_PRB2024,Yoshida_PRB2018}.
%

\begin{figure}[!t]
\begin{minipage}{0.9\hsize}
\begin{center}
\includegraphics[width=1\hsize,clip]{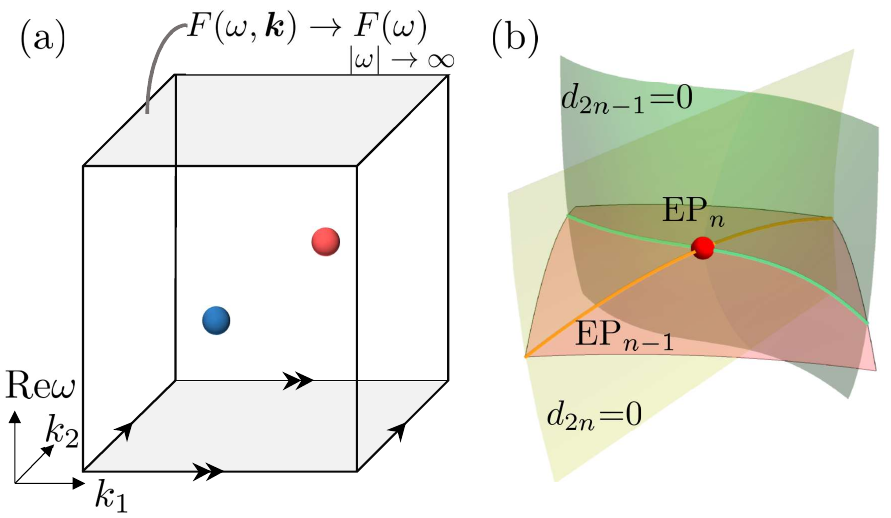}
\end{center}
\end{minipage}
\caption{
Illustration of the doubling theorem of nonlinear EP$n$s in $2n$-dimensional $\omega$-$\bm{k}$ space.
Panel (a) shows that an EP$n$ with $W_{2n-1}=1$ (red dot) is accompanied by another EP$n$ with $W_{2n-1}=-1$ (blue dot) due to the periodicity of the BZ and $\bm{k}$ independence of $F(\omega,\bm{k})$ for $|\omega|\to \infty$ [see Eqs.~\eqref{eq: W no symm} and \eqref{eq: Wtot=0 nosymm}]. 
In this panel, the axes of $\mathrm{Im}\omega$ and $k_j$ ($j=3,\ldots, 2n-2$) are omitted.
Panel (b) illustrates the winding structure of $\bm{d}(\omega,\bm{k})$ [Eq.~\eqref{eq: dvec nosymm}] in the vicinity of the EP$n$. 
The green (orange) surface separates the regions of positive and negative $d_{2n-1}$ ($d_{2n}$). 
The red surface denotes the region where $d_1=d_2=\ldots=d_{2n-2}=0$ holds, forming a manifold of EP$n'$ with $n'=n-1$. 
The vector $\bm{d}$ winds the $(2n-1)$-sphere in the vicinity of the EP$n$ emerging on the two-dimensional manifold of EP$n'$s with $n'=n-1$.
}
\label{fig: Overview}
\end{figure}

We begin by considering the codimension of nonlinear EP$n$s without symmetry in frequency-momentum ($\omega$-$\bm{k}$) space. An EP$n$ emerges at $\omega_0\in \mathbb{C}$ and $\bm{k}_0$ if and only if the following conditions are satisfied:
\begin{align}
\label{eq: partial^l det F}
\det F(\omega,\bm{k}_0)|_{\omega=\omega_0} &= 0, \nonumber \\
\partial_\omega \det F(\omega,\bm{k}_0)|_{\omega=\omega_0} &= 0, \nonumber \\
\vdots \nonumber \\
\partial^{n-1}_\omega \det F(\omega,\bm{k}_0)|_{\omega=\omega_0} &= 0,
\end{align}
and $\partial^n_{\omega} \det F(\omega,\bm{k}_0)|_{\omega=\omega_0} \neq 0$.
Since $\det F$ is a complex function,  the emergence of an EP$n$ requires satisfying $2n$ real-valued constraints.
Therefore, the codimension is $\delta_{\omega\text{-}\bm{k}}=2n$ in $D+2$-dimensional $\omega$-$\bm{k}$ space; i.e., EP$n$s emerge in the $D=2(n-1)$-dimensional BZ.
In passing, eigenvectors coalesce at $(\omega,\bm{k})=(\omega_0,\bm{k}_0)$ unless additional conditions are satisfied. 

The above codimension argument leads to FM topology of EP$n$s for systems with an arbitrary number of bands. We rewrite Eq.~\eqref{eq: partial^l det F} as $\bm{d}(\omega_0,\bm{k}_0)=\bm{0}$ with 
\begin{align}
\label{eq: dvec nosymm}
\bm{d}(\omega,\bm{k})
&=\Big(\mathrm{Re}[\det F],\mathrm{Im}[\det F],\nonumber \\
&\quad\quad \ldots,  \mathrm{Re}[\partial^{n-1}_\omega \det F],\mathrm{Im}[\partial^{n-1}_\omega \det F] \Big).
\end{align}
The vector $\bm{d}(\omega,\bm{k})$ is nonzero on the $M$-sphere $S^{M}$ with $M=\delta_{\omega\text{-}\bm{k}}-1$ enclosing the EP$n$ in $\omega$-$\bm{k}$ space. 
Thus, $\bm{d}(\omega,\bm{k})$ defines a map from $S^{M}$ in the $\omega$-$\bm{k}$ space to the unit vector $\hat{\bm{d}}(\omega,\bm{k})=\bm{d}(\omega,\bm{k})/\lVert\bm{d}(\omega,\bm{k})\rVert$ on $S^M$ whose topology in $\omega$-$\bm{k}$ space is classified by the homotopy group $\pi_{M}(S^M)=\mathbb{Z}$.

This topology is characterized by the FM winding number
\begin{align}
\label{eq: W no symm}
 W_{M} &= \frac{1}{A_{M}}
 \oint d^M\bm{p} \epsilon^{\mu_1\ldots\mu_{M+1}}f_{\mu_1\cdots\mu_{M+1}}
\end{align}
with
\begin{align}
f_{\mu_1\cdots\mu_{M+1}} &= \hat{d}_{\mu_1}\partial_1 \hat{d}_{\mu_2}\partial_2\hat{d}_{\mu_3}\cdots \partial_{M} \hat{d}_{\mu_{M+1}},
\end{align}
\begin{align}
A_M &= 2\pi^{\frac{M+1}{2}}\Big[\Gamma\big(\frac{M+1}{2}\big)\Big]^{-1}.
\end{align}
%
The integral is computed over $S^M$ in $\omega$-$\bm{k}$ space parameterized by $\bm{p}$. The symbol $\epsilon^{\mu_1\cdots\mu_{M+1}}$ denotes an anti-symmetric tensor satisfying $\epsilon^{12\cdots M+1}=1$. The derivative with respect to $p_i$ is denoted by $\partial_i$.
The area of the $M$-dimensional sphere $A_M$ is expressed in terms of the gamma function, which satisfies $\Gamma(z+1)=z\Gamma(z)$, $\Gamma(1)=1$, and $\Gamma(1/2)=\sqrt{\pi}$; 
its explicit form is given in Sec.~\ref{sec: A_M} of the Supplemental Material~\cite{Suppl}.

Notably, the above FM topological argument reveals the hierarchical structure of nonlinear EP$n$s. 
A manifold of EP$n$s emerges on a manifold of EP$n'$s whose dimension exceeds that of EP$n$s by $2(n-n')$ as illustrated in Fig.~\ref{fig: Overview}(b) for $n'=n-1$ in the $D$-dimensional BZ with $D=2n-2$. 
This structure follows directly from Eqs.~\eqref{eq: partial^l det F} and \eqref{eq: W no symm}.

\sectitle{Doubling of nonlinear EP$n$s without symmetry}
%
The FM winding number [see Eq.~\eqref{eq: W no symm}] leads to the doubling theorem of nonlinear EP$n$s. 
We consider EP$n$s in $2n$-dimensional $\omega$-$\bm{k}$ space [i.e., in the $D=2(n-1)$ dimensional BZ] for an $m$-band system ($m \geq n$). 
Because of the periodicity of the BZ, the total winding number $W_{\mathrm{tot},M}$ with $M=2n-1$ is determined by boundary terms at $|\omega|\to \infty$ (see Fig.~\ref{fig: Overview}). 
Since $\bm{d}(\omega,\bm{k})$ is $\bm{k}$ independent for $|\omega|\to \infty$, the integrant including momentum derivatives vanishes 
\begin{align}
\label{eq: Wtot=0 nosymm}
 W_{\mathrm{tot},M} & \propto \sum_{j}\oint_{S^{M}_j} d^M\bm{p}\, \epsilon^{\mu_1\ldots\mu_{M+1}}f_{\mu_1\cdots\mu_{M+1}} 
 \nonumber \\
 &\quad =\oint_{|\omega|\to\infty} d^M\bm{p}\, \epsilon^{\mu_1\ldots\mu_{M+1}}f_{\mu_1\cdots\mu_{M+1}}
=0.
\end{align}
This result implies that any EP$n$ with $W_{2n-1}=1$ is necessarily paired with a corresponding EP$n$ with $W_{2n-1}=-1$, thereby establishing the doubling theorem of EP$n$ for arbitrary $m$ ($m\geq n$), which had not been proved even in the linear case.

The FM winding number does not require any information about the band structure and is not affected by fake EP$n$s reported in a linear system~\cite{Yoshida_SciPostPhys2026}.
In addition, our FM winding number reproduces the doubling theorem of EP2s in the linear regime (see Sec.~\ref{sec: doubling linear EP2} of the Supplemental Material~\cite{Suppl}), which was previously obtained by using the discriminant of the characteristic polynomial~\cite{Yang_PRL2021}.

\sectitle{Cases with symmetry: an overview}
%
FM winding numbers also imply the doubling of nonlinear EP$n$s protected by symmetry.
%
\begin{table}[t]
    \centering
    \begin{tabular}{ccccl} 
\hline\hline
Symmetry                       & $\omega$              & $n$ & Codim. $\delta_{\omega\text{-}\bm{k}}$         & ~~$\bm{d}$~~                     \\ \hline 
none                           & $\mathbb{C}$          & e/o & $2n$                                    & Eq.~\eqref{eq: dvec nosymm}        \\
$PT$                           & $\mathbb{R}$          & e/o & $n$                                     & Eq.~\eqref{eq: dvec PT}($\ddagger$)            \\
\multirow{2}{*}{$CP$}          & \multirow{2}{*}{$0$}  & e   & $n$                                     & Eq.~\eqref{eq: dvec CP even N}     \\
                               &                       & o   & $n-1$                                   & Eq.~\eqref{eq: dvec CP odd N}      \\
\multirow{2}{*}{$PT$ and $CP$} & \multirow{2}{*}{$0$}  & e   & $n/2$                                   & Eq.~\eqref{eq: dvec PT+CP even N}  \\
                               &                       & o   & $(n-1)/2$                               & Eq.~\eqref{eq: dvec PT+CP odd N}   \\
chiral                         & $\ii \mathbb{R}$      & e/o & $n$                                     & Eq.~\eqref{eq: dvec PT}(*)     \\
$PT^\dagger$                   & $\mathbb{C}$          & e/o & $2n$                                    & Eq.~\eqref{eq: dvec nosymm}        \\
$CP^\dagger$                   & $\ii \mathbb{R}$          & e/o & $n$                                     & Eq.~\eqref{eq: dvec PT}(*)($\ddagger$)            \\
\multirow{2}{*}{sublattice}    & \multirow{2}{*}{$0$}  & e   & $n$                                     & Eq.~\eqref{eq: dvec CP even N}     \\
                               &                       & o   & $n-1$                                   & Eq.~\eqref{eq: dvec CP odd N}      \\
psH                            & $\mathbb{R}$          & e/o & $n$                                     & Eq.~\eqref{eq: dvec PT}        \\
\hline\hline
\end{tabular}
\caption{
The codimension $\delta_{\omega\text{-}\bm{k}}$ and the vector $\bm{d}$ for each case of symmetry, which specify the explicit form of the FM winding number in Eq.~\eqref{eq: W no symm}.
The first column denotes symmetry, where pseudo-Hermiticity is denoted by ``psH".
The second column indicates the frequency $\omega$ where the symmetries are closed.
The third column indicates the parity of $n$; ``e" (``o") denotes even (odd) $n$, and ``e/o" denotes even or odd $n$. 
We consider symmetry operations that square to $1$.
The codimension and degrees of freedom of $\omega$ indicates that nonlinear EP$n$s emerge in the $D$-dimensional BZ with $D=\delta_{\omega\text{-}\bm{k}}-\mathrm{dim}\omega$ ($\mathrm{dim}\omega=2$ 
for $\omega \in \mathbb{C}$, $\mathrm{dim}\omega=1$ for $\omega \in \mathbb{R}\cup \ii \mathbb{R}$, and $\mathrm{dim}\omega=0$ for $\omega=0$). 
For the cases marked with the symbol ``(*)", $\bm{d}$ is defined by replacing $F\to \tilde{F}:=-\ii F$ and $\omega\to \tilde{\omega}:=-\ii\omega$.
In the cases marked with the symbol ``($\ddagger$)", EP2s are characterized by a $\mathbb{Z}$ invariant rather than the previously proposed $\mathbb{Z}_2$ invariant in the linear limit, revealing the stability of these EP2s that was previously unrecognized.
}
    \label{tab: delta d maintxt}
\end{table}
%
%
%
We consider $PT$, $CP$, and chiral symmetries
\begin{align}
\label{eq: PT}
U_{\mathrm{PT}} F^*(\omega,\bm{k}) U^{-1}_{\mathrm{PT}} &= F(\omega^*,\bm{k}),
\\
\label{eq: CP}
U_{\mathrm{CP}} F^T(\omega,\bm{k}) U^{-1}_{\mathrm{CP}} &= -F(-\omega,\bm{k}),
\\
\label{eq: chiral}
U_{\Gamma} F^\dagger(\omega,\bm{k}) U^{-1}_{\Gamma} &= -F(-\omega^*,\bm{k}).
\end{align}
We assume that the unitary matrices satisfy $U_{\mathrm{PT}}U^*_{\mathrm{PT}}=1$, $U_{\mathrm{CP}}U^*_{\mathrm{CP}}=1$, and  $U^2_{\Gamma}=1$, respectively.
These symmetries impose additional conditions on $\omega$ and $\det F(\omega,\bm{k})$. 
Taking these conditions into account, we obtain the codimension $\delta_{\omega\text{-}\bm{k}}$ and $\bm{d}(\omega,\bm{k})$ (see Table~\ref{tab: delta d maintxt}) as discussed below. 

Table~\ref{tab: delta d maintxt} shows that symmetry-protected EP$n$s emerge in $\delta_{\omega\text{-}\bm{k}}$-dimensional $\omega$-$\bm{k}$ space, i.e., the $D$-dimensional BZ with $D=\delta_{\omega\text{-}\bm{k}}-\mathrm{dim}\omega$ and $\mathrm{dim}\omega$ which takes $\mathrm{dim}\omega=2$, $1$, and $0$ for $\omega \in \mathbb{C}$, $\omega \in \mathbb{R}\cup \ii \mathbb{R}$, and $\omega=0$, respectively.
As is the case without symmetry, EP$n$s possess FM topology classified by homotopy group $\pi_M(S^M)=\mathbb{Z}$ with $M=\delta_{\omega\text{-}\bm{k}}-1$. The corresponding topological invariant is the winding number in Eq.~\eqref{eq: W no symm} with $\bm{d}(\omega,\bm{k})$ listed in Table~\ref{tab: delta d maintxt}. 
Due to the periodicity of the BZ and the $\bm{k}$ independence of $\det F$ for $|\omega|\to \infty$, the total winding number of symmetry-protected EP$n$s vanishes [see Eq.~\eqref{eq: Wtot=0 nosymm}].
Thus, a symmetry-protected EP$n$ with $W_{M}=1$ for $M=\delta_{\omega\text{-}\bm{k}}-1$ is accompanied by another one with $W_{M}=-1$, thereby establishing the doubling theorem.

The above argument also extends to cases with $PT^\dagger$ symmetry, $CP^\dagger$ symmetry, sublattice symmetry, and pseudo-Hermiticity (see Table~\ref{tab: delta d maintxt}), as discussed in Sec.~\ref{sec: other symm} of the Supplemental Material~\cite{Suppl}.

\sectitle{Codimension under $PT$ symmetry}
Because $PT$ transformation maps $\omega$ to $\omega^*$, the constraint in Eq.~\eqref{eq: PT} is closed for $\omega\in \mathbb{R}$. Thus, we focus on symmetry-protected EP$n$s that emerge on the real axis of frequency where $\det F(\omega,\bm{k})$ is real.

In this case, the conditions of symmetry-protected EP$n$s [see Eq.~\eqref{eq: partial^l det F}]
is written as $n$ real constraints $\bm{d}(\omega_0,\bm{k}_0)=\bm{0}$ with
\begin{align}
\label{eq: dvec PT}
\bm{d}(\omega,\bm{k}) &=
\Big(
\det F,
\partial_\omega \det F,
\ldots,\partial^{n-1}_\omega \det F
\Big).
\end{align}
Thus, their codimension is $\delta_{\omega\text{-}\bm{k}}=n$, indicating that symmetry-protected EP$n$s emerge in the $D=(n-1)$-dimensional BZ. 
The above vector $\bm{d}(\bm{k})$ reveals FM topology of symmetry-protected EP$n$s characterized by the winding number in Eq.~\eqref{eq: W no symm}, thereby leading to the doubling theorem.

In passing, the condition $\bm{d}(\omega_0,\bm{k})=\bm{0}$ in FM space reveals the hierarchical structure of EP$n$s with $PT$ symmetry.
A manifold of EP$n$s emerges on a manifold of EP$n'$s whose dimension exceeds that of the EP$n$s by $(n-n')$ [see Eq.~\eqref{eq: dvec PT}].

In the linear limit, it is known that a $\mathbb{Z}_2$ invariant characterizes EP2s with $PT$ symmetry, whereas our topological invariant takes arbitrary integer values. Our numerical analysis demonstrates the presence of EP2s with $PT$ symmetry beyond the previously discussed $\mathbb{Z}_2$ topology (for more details, see Sec.~\ref{sec: ZvsZ2 EP2 PT} of the Supplemental Material~\cite{Suppl}).

\sectitle{Codimension under $CP$ symmetry}
%
%
Because $CP$ transformation maps $\omega$ to $-\omega$, the constraint in Eq.~\eqref{eq: CP} is closed for $\omega=0$, indicating that $CP$ symmetry protects EP$n$s only at $\omega_0=0$.
In addition, Eq.~\eqref{eq: CP} leads to 
\begin{align}
\label{eq: CP partial^l detF}
\partial^{l}_\omega \det F(\omega,\bm{k})\Big|_{\omega=\omega_0} &= 0
\end{align}
for $l$ and $N$ of opposite parity.
Here, $N$ denotes the size of the matrix $F$.
Thus, one of the conditions of EP$n$s, $\partial^n_\omega \det F(\omega,\bm{k}_0)|_{\omega=\omega_0}\neq 0$, is compatible only for $n=2s+2$ when $N$ is even, and for $n=2s+3$ when $N$ is odd ($s=0,1,2,\ldots$).
The remaining conditions for symmetry-protected EP$n$s lead to $\bm{d}(\omega_0,\bm{k}_0)=\bm{0}$ with
\begin{subequations}
\label{eq: dvec CP}
\begin{align}
\label{eq: dvec CP even N}
\bm{d}(\omega_0,\bm{k}) &=
\Big(
\mathrm{Re}[\det F],
\mathrm{Im}[\det F],
\mathrm{Re}[\partial^2_{\omega_0} \det F],
\nonumber \\
&\fivequad \quad\quad
\ldots,\partial^{2s}_{\omega_0} \mathrm{Im}[\det F]\Big)
\end{align}
for $n=2s+2$ (i.e., even $N$) and 
\begin{align}
\label{eq: dvec CP odd N}
\bm{d}(\omega_0,\bm{k}) &=
\Big(
\mathrm{Re}[\partial_{\omega_0} \det F],
\mathrm{Im}[\partial_{\omega_0} \det F],
\nonumber \\
&
\quad\quad
\mathrm{Re}[\partial^3_{\omega_0} \det F],
\ldots,\mathrm{Im}[\partial^{2s+1}_{\omega_0} \det F]
\Big)
\end{align}
\end{subequations}
for $n=2s+3$ (i.e., odd $N$). 
Here, we have used a simplified notation, 
$\partial^l_{\omega_0}\mathrm{det}F:= \partial^l_{\omega}\mathrm{det}F(\omega,\bm{k})|_{\omega=\omega_0}$.
We stress that in the above definition, $\omega$ is fixed at $\omega_0=0$ in contrast to that in Eq.~\eqref{eq: dvec PT}. 
Equation~\eqref{eq: dvec CP} indicates that the codimension of symmetry-protected EP$n$s is $\delta_{\omega\text{-}\bm{k}}=n$ [$\delta_{\omega\text{-}\bm{k}}=n-1$] for even [odd] $n$.
The vector $\bm{d}(\bm{k})$ reveals FM topology of symmetry-protected EP$n$s characterized by the winding number in Eq.~\eqref{eq: W no symm}, thereby establishing the doubling theorem.

In passing, the condition $\bm{d}(\omega_0,\bm{k})=\bm{0}$ in FM space reveals the hierarchical structure of EP$n$s with $CP$ symmetry.
A manifold of EP$n$s emerges on a manifold of EP$n'$s whose dimension exceeds that of EP$n$s by $n-n'$, provided that $n$ and $n'$ have the same parity [see Eq.~\eqref{eq: dvec CP}].

\sectitle{Codimension under $PT$ and $CP$ symmetries}
Because these symmetries are closed only at $\omega=0$, we set $\omega_0=0$. Due to $PT$ symmetry, $\mathrm{det}F(\omega_0,\bm{k})$ is real. In addition, due to $CP$ symmetry, Eq.~\eqref{eq: CP partial^l detF} holds, indicating that symmetry-protected EP$n$s with $n=2s+2$ [$n=2s+3$] are compatible for even $N$ [odd $N$] ($s=0,1,2,\ldots$). 
This fact leads to the constraints of symmetry-protected EP$n$s, $\bm{d}(\omega_0,\bm{k})=\bm{0}$ with 
\begin{subequations}    
\label{eq: dvec PT+CP}
\begin{align}
\label{eq: dvec PT+CP even N}
\bm{d}(\omega_0,\bm{k}) &=
\Big(
\det F,
\partial^2_{\omega_0} \det F,
\ldots,
\partial^{2s}_{\omega_0}\det F\Big)
\end{align}
for $n=2s+2$ and 
\begin{align}
\label{eq: dvec PT+CP odd N}
\bm{d}(\omega_0,\bm{k}) &=
\Big(
\partial_{\omega_0}\det F,
\partial^3_{\omega_0} \det F,
\ldots,\partial^{2s+1}_{\omega_0} \det F
\Big)
\end{align}
\end{subequations}    
for $n=2s+3$.
We recall that $\omega$ is fixed to $\omega_0=0$ and that we have used a simplified notation, $\partial^l_{\omega_0}\mathrm{det}F
:=
\partial^l_{\omega}\mathrm{det}F(\omega,\bm{k})|_{\omega=\omega_0}$. 
Equation~\eqref{eq: dvec PT+CP} indicates that the codimension of symmetry-protected EP$n$s is $\delta_{\omega\text{-}
\bm{k}}=n/2$ [$\delta_{\omega\text{-}\bm{k}}=(n-1)/2$] for $n=2s+2$ ($n=2s+3$) with $s=0,1,2,\ldots$.  
The above vector $\bm{d}(\bm{k})$ reveals FM topology of symmetry-protected EP$n$s with $n \geq 4 $ characterized by the winding number in Eq.~\eqref{eq: W no symm} which leads to the doubling theorem. 

For $n=2,3$, symmetry-protected EP$n$s are characterized by the $\mathbb{Z}_2$ invariant $\nu$
\begin{align}
\label{eq: sgn d_1}
(-1)^\nu &=\prod_{k\in\{k_{+0},k_{-0} \}} \mathrm{sgn}\big[d_1(\omega_0,k)\big]
\end{align}
with $d_1(\omega_0,k)=\mathrm{det}F(\omega_0,k)$ [$d_1(\omega_0,k)=\partial_{\omega_0}\det F(\omega_0,k)$] for $n=2$ [$n=3$]. 
The momentum slightly displaced from the EP$n$ is denoted by $k_{\pm 0}:=k_0\pm\delta k$ with $\delta k >0$. This topological invariant leads to the doubling theorem due to the periodicity of the BZ. 

In passing, the condition $\bm{d}(\omega_0,\bm{k})=\bm{0}$ in FM space reveals the hierarchical structure of EP$n$s with $PT$ and $CP$ symmetries.
A manifold of EP$n$s emerges on a manifold of EP$n'$s whose dimension exceeds that of EP$n$s by $(n-n')/2$, provided that $n$ and $n'$ have the same parity [see Eq.~\eqref{eq: dvec PT+CP}].

\sectitle{Codimension under chiral symmetry}
%
Chiral symmetry~\eqref{eq: chiral} is closed for $\omega\in \ii \mathbb{R}$. Thus, we focus on symmetry-protected EP$n$s that emerge on the imaginary axis. In addition, the chiral symmetry is rewritten as
\begin{align}
 U_{\Gamma} \tilde{F}^\dagger(\tilde{\omega},\bm{k}) U^\dagger_{\Gamma} &= \tilde{F}(\tilde{\omega}^*,\bm{k})
\end{align}
with $\tilde{F}(\tilde{\omega},\bm{k})=-\ii F(\ii \tilde{\omega},\bm{k})$ and $\tilde{\omega}=-\ii \omega$~\cite{Kawabata_NatComm2019,Kawabata_PRX2019}.
Because the transpose of an arbitrary matrix does not affect its determinant, the problem is reduced to the case with $PT$ symmetry. 
Thus, the codimension is $\delta_{\omega\text{-}\bm{k}}=n$. The vector $\bm{d}(\tilde{\omega},\bm{k})$ is obtained in the same way as for deriving Eq.~\eqref{eq: dvec PT} except that $F$ and $\omega$ are replaced by $\tilde{F}$ and $\tilde{\omega}$, respectively.

\sectitle{
Application to a toy model with $PT$ symmetry
}
%
\begin{figure}[!tb]
\begin{minipage}{0.95\hsize}
\begin{center}
\includegraphics[width=1\hsize,clip]{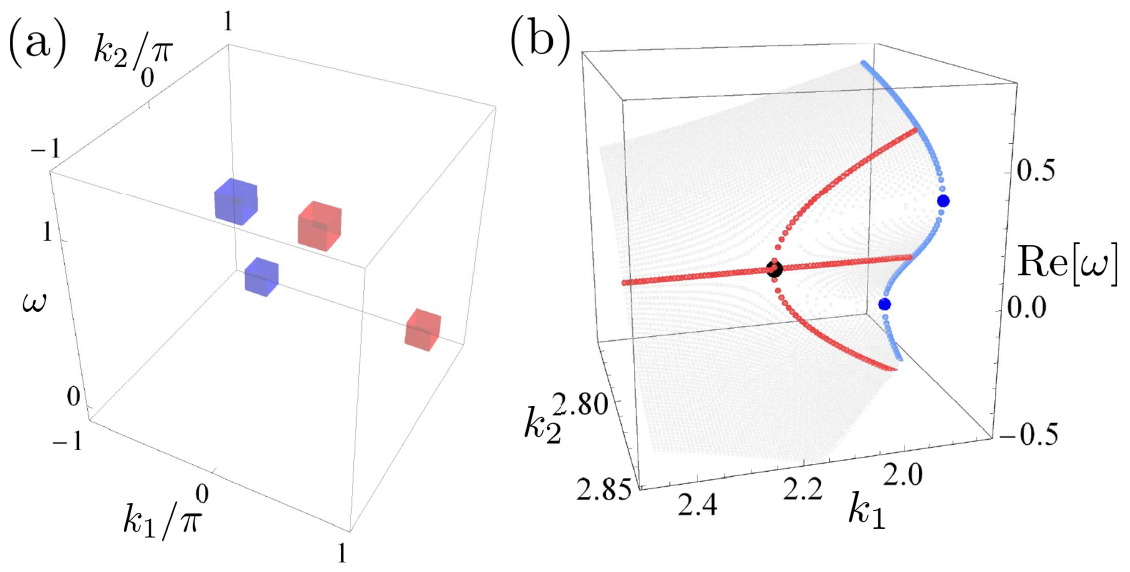}
\end{center}
\end{minipage}
\caption{
(a) Winding number $W_2$ for each mesh. The winding number is $W_2=1$ ($W_2=-1$) on the surface of red (blue) cuboids. Here, we take a $10\times10\times10$ mesh. The winding number is computed based on the method proposed in Ref.~\cite{Fukui_JPSJ2005}. For the computation of $W_2$, each cuboid is further subdivided into a $10\times10\times 10$ mesh.
(b) Eigenvalues $\omega\in\mathbb{R}$ as functions of $\bm{k}=(k_1,k_2)$. The data for $k_2=2.80176$ ($k_1=2.15984$) are represented by red (blue) points. The others are represented by gray points. The blue (black) point denotes an EP2 (EP3).
Data are obtained for $(\delta_1,\delta_2)=(0.01,1)$.
}
\label{fig: EP3inPT}
\end{figure}
%
Numerical analysis of a two-dimensional toy model with $PT$ symmetry demonstrates the doubling of nonlinear EP$3$s whose codimension is $\delta_{\omega\text{-}\bm{k}}=3$. 
The model is described by
$F(\omega,\bm{k}) = \omega\1 -H_0(\bm{k})-\Sigma(\omega)$ with a $4\times 4$ matrix $H_0(\bm{k})$ exhibiting EP$3$ and $\Sigma(\omega)$ including nonlinearity of $\omega$ (for explicit forms of $H_0(\bm{k})$ and $\Sigma(\omega)$, see Sec.~\ref{sec: detail toy EP3PT} of the Supplemental Material~\cite{Suppl}).

The EP$3$s are characterized by the winding number $W_2$  with vector $\bm{d}$ [see Eqs.~\eqref{eq: W no symm} and \eqref{eq: dvec PT}]. 
Figure~\ref{fig: EP3inPT}(a) displays $W_2$ for each point in the $\omega$-$\bm{k}$ space. This figure indicates that the points of $W_2=1$ are accompanied by the ones of $W_2=-1$. In addition, the point with $W_2=1$ captures a triple root [see Fig.~\ref{fig: EP3inPT}(b)]. These data numerically confirm the doubling theorem of EP$3$s with $PT$ symmetry.

Here, we note that the topology of arbitrary $\bm{d}$ is rephrased in terms of a Hermitian Hamiltonian: $H_{\bm{d}}(\bm{q})=\sum_i d_i \gamma_i$ with Hermitian matrices $\gamma$'s satisfying $\gamma_i\gamma_j + \gamma_j\gamma_i=2\delta_{ij}$ and $\bm{q}=(\omega,k_1,\ldots,k_D)^T$  (for more details, see Sec.~\ref{sec: Hd} of the Supplemental Material~\cite{Suppl}). This mapping allows us to compute $W_2$ by employing the method proposed in Refs.~\cite{Fukui_JPSJ2005}.

\sectitle{
Topology of an EP3 in coupled resonators with nonlinearity of eigenvectors
}
FM topology protects an EP3 in a class of coupled resonators with saturable gain~\cite{Bai_PRL2023,Bai_NatlSciRev2022,Fang_PRB2025} described by $H_{|\psi\rangle}|\psi\rangle =\omega |\psi\rangle$ with 
\begin{align}
H_{|\psi\rangle} &=
\left(
\begin{array}{cc}
\omega_\mathrm{A}+\ii g_\mathrm{A}(|\psi_\mathrm{A}|) & \kappa_1  \\
\kappa_1 & \omega_{\mathrm{B}}-\ii l_{\mathrm{B}}
\end{array}
\right).
\end{align}
Here, $g_{\mathrm{A}}(|\psi_A|)$ denotes a nonlinear saturable gain. The resonant frequencies of each resonator are denoted by $\omega_{\mathrm{A}}$ and $\omega_{\mathrm{B}}$, respectively. The linear loss is denoted by $l_{\mathrm{B}}$.
After reaching the stable state, the eigenvalues $\omega$ satisfy $p(\omega)=0$ with
$p(\omega) = (\omega-\omega_{\mathrm{A}})\big[(\omega-\omega_{\mathrm{B}})^2 +l^2_{\mathrm{B}}\big]-\kappa^2_1(\omega-\omega_{\mathrm{B}})$.

For $\omega_{\mathrm{A}}=\omega_{\mathrm{B}}=0$, eigenvalues are written as $\omega=0$ and $\omega=\pm\sqrt{\kappa^2_1-l^2_{\mathrm{B}}}$, indicating that the system hosts an EP3 despite the $2\times 2$ matrix $H_{|\psi\rangle}$.
Our approach reveals the topological protection of these EP3s.
In the vicinity of $(\omega,\omega_\mathrm{A},l_{\mathrm{B}})=(0,0,\kappa_1)$ with $\kappa_1>0$,
$\bm{d}(\delta \omega,\delta \omega_{\mathrm{A}},\delta l_{\mathrm{B}})$ 
for $|\delta \omega|,|\delta \omega_{\mathrm{A}}|,|\delta l_\mathrm{B}|\ll \kappa_1$
is given by
$\bm{d}^T =
\Big(-\kappa^2_1\delta \omega_{\mathrm{A}},2\kappa_1\delta l_{\mathrm{B}},6\delta \omega-2\delta \omega_{\mathrm{A}}\Big)$
from which we obtain the winding number $W_2=-1$.
A similar analysis applies to an EP$3$ in Hermitian nonlinear coupled resonators with Kerr nonlinearity~\cite{Fang_PRB2025}. 

\sectitle{Nonlinear Kramers pairs}
While we have focused on the cases of $U_{\mathrm{PT}}U^*_{\mathrm{PT}}=1$ so far, the doubling theorem for cases of $U_{\mathrm{PT}}U^*_{\mathrm{PT}}=-1$ remains an open question due to nonlinear Kramers pairs.
If $|\psi_n\rangle$ is an eigenstate with eigenvalue $\omega_n$, then $U_{\mathrm{PT}}\mathcal{K}|\psi_{n}\rangle$ is also an eigenstate with eigenvalue $\omega^*_n$ where $\mathcal{K}$ denotes the complex conjugate operator.
This degeneracy prevents a sign change of $\det F(\omega,\bm{k})$ in the vicinity of EP$n$s.
A similar argument also holds for $CP^\dagger$ symmetry since it reduces to $PT$ symmetry.
%

\sectitle{Conclusion}
%
We have established the doubling theorem of EP$n$s for $m$-band systems with nonlinearity of frequency. The introduced FM winding numbers reveal a global topological structure of EP$n$s over the entire BZ and underlie the doubling theorem both in the absence and presence of symmetry.
In the linear limit, our framework suggests $\mathbb{Z}$ topology for $PT$-symmetric EP2s beyond the previously discussed $\mathbb{Z}_2$ topology.
The same framework directly covers dispersive metamaterials and quantum quasiparticles with finite lifetime. It can be extended to a class of coupled resonators with nonlinearity of eigenvectors whose spectrum is determined by a nonlinear scalar equation for the frequency.
The extension to cases with nonlinear Kramers pairs remains an open question.

\sectitle{Acknowledgments}
The author thanks Emil J. Bergholtz and Tom\'a\v{s} Bzdu\v{s}ek for the previous collaboration~\cite{Yoshida_SciPostPhys2026}, which provided part of the motivation for this work.
This work is supported by JSPS KAKENHI Grant Nos.~JP21K13850, JP23KK0247, JP25K07152, and JP25H02136 as well as JSPS Bilateral Program No.~JPJSBP120249925.

%


\clearpage

\renewcommand{\thesection}{S\arabic{section}}
\renewcommand{\theequation}{S\arabic{equation}}
\setcounter{equation}{0}
\renewcommand{\thefigure}{S\arabic{figure}}
\setcounter{figure}{0}
\renewcommand{\thetable}{S\arabic{table}}
\setcounter{table}{0}
\makeatletter
\c@secnumdepth = 2
\makeatother

\twocolumngrid
\begin{center}
 {\large \textmd{Supplemental Material:} \\[0.3em]
 {\bfseries 
  Nonlinear Frequency-Momentum Topology and Doubling of Multifold Exceptional Points
 }
 }
\end{center}

\setcounter{page}{1}

\section{
Area of the $M$-dimensional sphere
}
\label{sec: A_M}

For an integer $n$, the gamma function is written as
\begin{align}
\label{eq: Gamma(n)}
\Gamma(n) &=(n-1)!, \\
\label{eq: Gamma(n+1/2)}
\Gamma\big(n+\frac{1}{2}\big) &= \frac{(2n)!}{2^{2n}n!}\sqrt{\pi}.
\end{align}
Thus, the area of the $(2M+1)$-dimensional sphere is 
\begin{align}
\label{eq: A_(2M+1)}
A_{2M+1} &=\frac{2\pi^{M+1}}{M!},
\end{align}
and that of the $2M$-dimensional sphere is 
\begin{align}
\label{eq: A_(2M)}
A_{2M} &=\frac{2^{2M+1}\pi^M M!}{(2M)!}.
\end{align}

\section{Frequency-momentum winding number in the linear limit}
\label{sec: doubling linear EP2}

In the linear limit, the FM winding number coincides with the discriminant winding number~\cite{Yang_PRL2021} and thus reproduces the doubling theorem of EP$2$s.

We see the coincidence of these two topological invariants. We consider a toy model given by a $2\times 2$ matrix
\begin{align}
 H(\bm{k}) &=
 \pmat{ 0  & 1 \\
  z_{\eta}(\bm{k}) & 0
 }
\end{align}
with an integer $\eta$ and $z_{\eta}(\bm{k})=(k_1+ \ii k_2)^\eta$ for $\eta>0$, $z_{\eta}(\bm{k})=(k_1- \ii k_2)^{|\eta|}$ for $\eta<0$ and $z_{0}(\bm{k})=k^2_1+k^2_2$ for $\eta=0$. This model hosts an EP$2$ at $\omega_0=0$ and $\bm{k}_0=\bm{0}$ characterized by the discriminant winding number $W_{\mathrm{D}}=\eta$~\cite{Yang_PRL2021,Yoshida_PRR2025}.

The FM winding number $W_3$ can be computed from 
\begin{align}
 & W_3 = \sum_{(\omega,\bm{k})\in\{(0,\bm{k}_l)\}} \mathrm{sgn}\Big( \mathrm{det}[J(\omega,\bm{k})]
 \Big)
\end{align}
with
\begin{align} 
 J(\omega,\bm{k}) &= \frac{\partial (d_1,d_2,d_3,d_4)}{\partial(\omega_\mathrm{R},\omega_\mathrm{I},k_1,k_2)},
\end{align}
\begin{align} 
 d_1+\ii d_2 &= \omega^2-z(\bm{k}), \\
 d_3+\ii d_4 &= 2\omega,
\end{align}
and $\omega=\omega_{\mathrm{R}}+\ii \omega_{\mathrm{I}}$ ($\omega_{\mathrm{R}},\omega_{\mathrm{I}}\in \mathbb{R}$). For $\eta>0$ [$\eta<0$], $\bm{k}_l=(k_{1l},k_{2l})$ satisfies $k_{1l}+\ii k_{2l}=e^{\frac{2\ii \pi l}{\eta}}$ [$k_{1l}-\ii k_{2l}=e^{\frac{2\ii \pi l}{|\eta|}}$] with $l=0,1,\ldots,\eta-1$.
Here, $\mathrm{sgn}(x)$ takes $1$ ($-1$) for $x>0$ ($x<0$). At the points $(0,\bm{k}_l)$, the normalized vector satisfies $\bm{d}/\lVert\bm{d}\rVert=(-1,0,0,0)$. 
By computing the Jacobian, we have $\mathrm{det}J(0,\bm{k}_l)=4\eta^2\mathrm{sgn}(\eta)$ for $\eta\neq0$ which indicates that the FM winding number is $W_3=\eta$. For $\eta=0$, the FM winding number vanishes $W_3=0$ because $d_2=0$ for arbitrary $(\omega,\bm{k})$. Because these two topological invariants are independent of the details of $H(\bm{k})$, this coincidence holds for an arbitrary model.

\section{
Other cases of symmetry
}
\label{sec: other symm}

FM topology can also be extended to cases with $PT^\dagger$, $CP^\dagger$, sublattice symmetries, and pseudo-Hermiticity (ps H)
\begin{align}
\label{eq: PT dag}
U_{\mathrm{PT}^\dagger} F^T(\omega,\bm{k}) U^{-1}_{\mathrm{PT}^\dagger} &= F(\omega,\bm{k}),
\\
\label{eq: CP dag}
U_{\mathrm{CP}^\dagger} F^*(\omega,\bm{k}) U^{-1}_{\mathrm{CP}^\dagger} &= -F(-\omega^*,\bm{k}),
\\
\label{eq: sublatt}
U_{\mathrm{S}} F(\omega,\bm{k}) U^{-1}_{\mathrm{S}} &= -F(-\omega,\bm{k}),
\\
\label{eq: psH}
U_{\mathrm{psH}} F^\dagger(\omega,\bm{k}) U^{-1}_{\mathrm{psH}} &= F(\omega^*,\bm{k}).
\end{align}
%
The codimension $\delta_{\omega\text{-}\bm{k}}$ and the vector $\bm{d}$ are summarized in Table~\ref{tab: delta d maintxt}.
As discussed below, these results are obtained by noting that the transpose of a matrix does not change its determinant $\det F^T=\det F$, which reduces the problem to cases with symmetries defined in Eqs.~\eqref{eq: PT}, \eqref{eq: CP}, and \eqref{eq: chiral}.

\subsection{
Codimension under $PT^\dagger$ symmetry
}
\label{sec: PT^dag}

$PT^\dagger$ symmetry is closed for arbitrary $\omega\in \mathbb{C}$. In addition, the transpose of an arbitrary matrix does not affect its determinant. 
Thus, the codimension and topology of EP$n$s are the same as those in the case with no symmetry.

\subsection{
Codimension under $CP^\dagger$ symmetry
}
\label{sec: CP^dag}

$CP^\dagger$ symmetry is closed for arbitrary $\omega\in \ii \mathbb{R}$. 
In addition, $CP^\dagger$ symmetry is rewritten as
\begin{align}
 U_{CP^\dagger} \tilde{F}^*(\tilde{\omega},\bm{k}) U^\dagger_{CP^\dagger} &= \tilde{F}(\tilde{\omega}^*,\bm{k})
\end{align}
with $\tilde{F}(\tilde{\omega},\bm{k})=-\ii F(\ii \tilde{\omega},\bm{k})$ and $\tilde{\omega}=-\ii \omega$.
The above constraint is nothing but Eq.~\eqref{eq: PT}. Thus, the codimension is $\delta_{\omega\text{-}\bm{k}}=n$. The vector $\bm{d}(\tilde{\omega},\bm{k})$ is obtained in the same way as Eq.~\eqref{eq: dvec PT} 
with $F$ and $\omega$ replaced by $\tilde{F}$ and $\tilde{\omega}$.

\subsection{
Codimension under sublattice symmetry
}
\label{sec: sublattice}
Sublattice symmetry is closed for $\omega=0$. In addition, the transpose of an arbitrary matrix does not affect its determinant. Thus, the codimension and topology of EP$n$s are the same as those in the case with $CP$ symmetry.

\subsection{
Codimension under pseudo-Hermiticity
}
\label{sec: psH}

Pseudo-Hermiticity is closed for $\omega\in\mathbb{R}$. In addition, the transpose of an arbitrary matrix does not affect its determinant. Thus, the codimension and topology of EP$n$s are the same as those in the case with $PT$ symmetry.

\section{
EP2 with $PT$ symmetry
}
\label{sec: ZvsZ2 EP2 PT}

In the linear limit, our winding number $W_1$ in Eq.~\eqref{eq: W no symm} with $\bm{d}=(\det F,\partial_\omega \det F)$ reveals the robustness of EP2s with $PT$ symmetry, which was not understood from previous $\mathbb{Z}_2$ invariants.

For concreteness, we consider toy models
\begin{align}
H_{4\times 4} &=
\left(
\begin{array}{cccc}
0 & 1& 0        & 0  \\
k & 0& V        & 0  \\
0 & V& 0        & 1  \\
0 & 0& k-\kappa & 0
\end{array}
\right)
\end{align}
and 
\begin{align}
H_{2\times 2} &=
\left(
\begin{array}{cc}
0 & 1  \\
m_0-\sin k & 0  \\
\end{array}
\right)
\end{align}
with real values $V$, $\kappa$, and $m_0$.

\begin{figure}[!b]
\begin{minipage}{0.95\hsize}
\begin{center}
\includegraphics[width=1\hsize,clip]{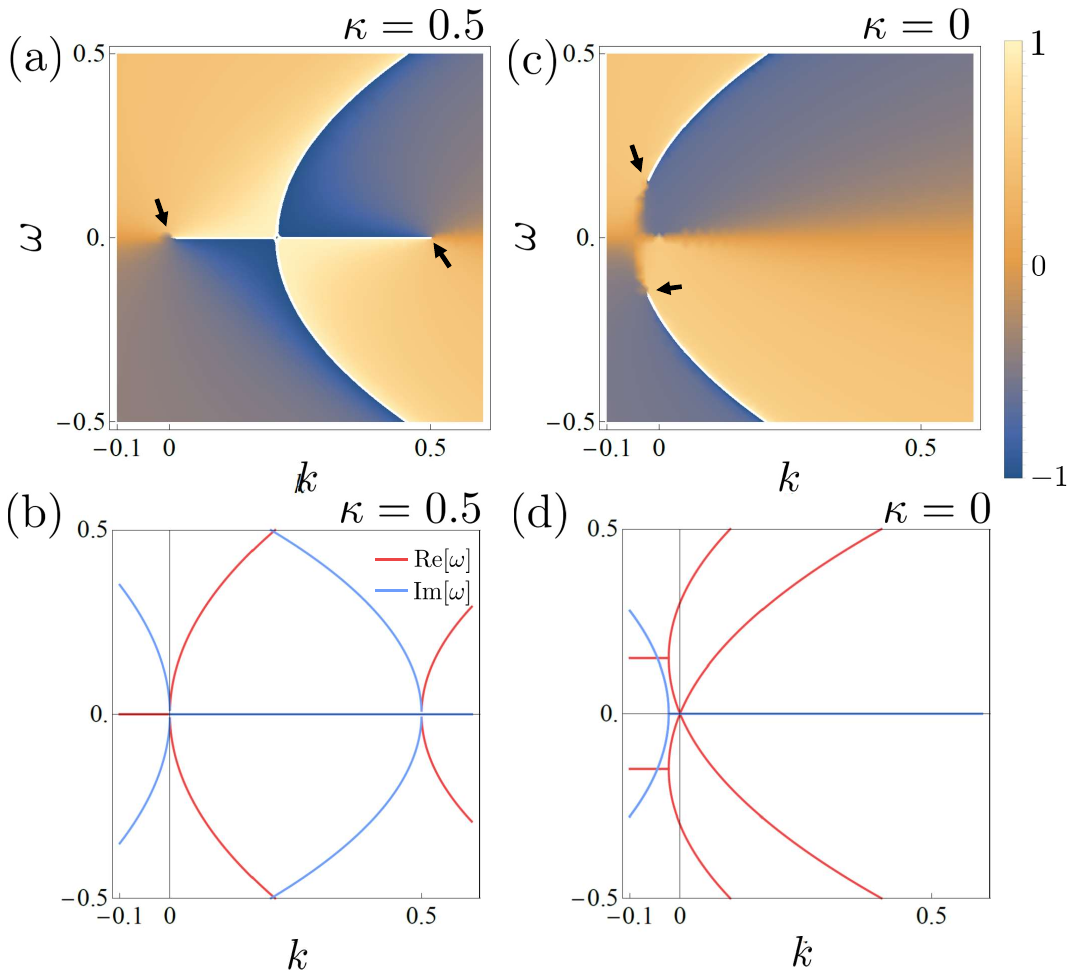}
\end{center}
\end{minipage}
\caption{
(a) [(c)]: Color plot of the argument $\mathrm{Arg}[d_1+\ii d_2]/\pi$ as a function of $k$ and $\omega$ ($\omega\in\mathbb{R}$) of the Hamiltonian $H_{4\times4}$ for $\kappa=0.5$ [$\kappa=0$]. In these panels, the FM winding number takes the value $W_1=-1$ along the path enclosing points denoted by black arrows.
(b) [(d)]: Eigenvalues as functions of $k$ of the Hamiltonian $H_{4\times4}$ for $\kappa=0.5$ [$\kappa=0$]. The real and imaginary parts of eigenvalues are represented by red and blue lines. 
The data are obtained for $V=0.3$.
}
\label{fig: EP2wPT 4x4}
\end{figure}

\begin{figure}[!tb]
\begin{minipage}{0.95\hsize}
\begin{center}
\includegraphics[width=1\hsize,clip]{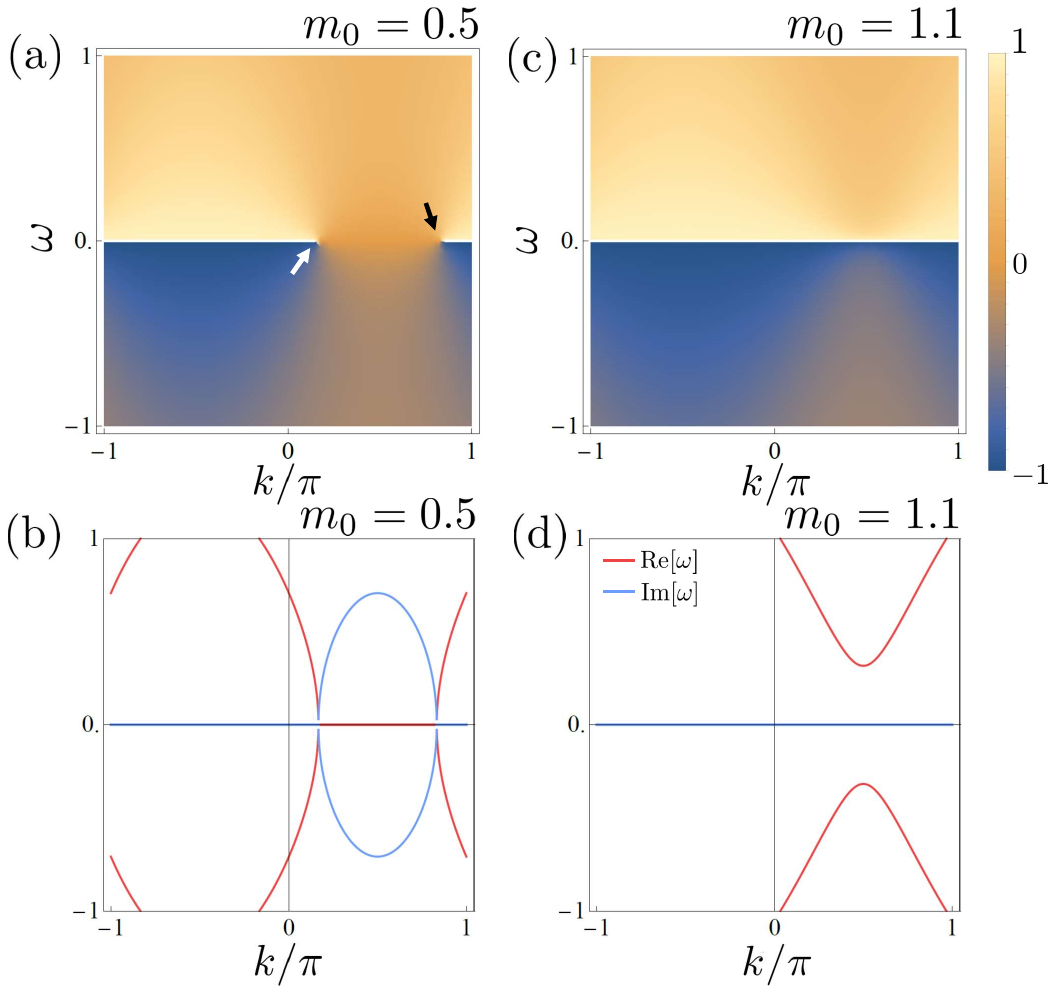}
\end{center}
\end{minipage}
\caption{
(a) [(c)]: Color plot of the argument $\mathrm{Arg}[d_1+\ii d_2]/\pi$ as a function of $k$ and $\omega$ ($\omega\in\mathbb{R}$) of the Hamiltonian $H_{2\times2}$ for $m_0=0.5$ [$m_0=1.1$]. In these panels, the FM winding number takes the value $W_1=-1$ ($W_1=1$) along the path enclosing points denoted by the black (white) arrow.
(b) [(d)]: Eigenvalues as functions of $k$ of the Hamiltonian $H_{2\times2}$ for $m_0=0.5$ [$m_0=1.1$]. The real and imaginary parts of eigenvalues are represented by red and blue lines. 
}
\label{fig: EP2wPT 2x2}
\end{figure}

The winding number and the band structure of $H_{4\times 4}$ are displayed in Fig.~\ref{fig: EP2wPT 4x4}.
Figures~\ref{fig: EP2wPT 4x4}(a) and \ref{fig: EP2wPT 4x4}(b) indicate the presence of two EP2s characterized by $W_1=-1$ for $\kappa=0.5$. As $\kappa$ decreases, these two EP2s approach each other. 
Notably, however, they do not annihilate each other because they have the same winding number.
This behavior is in sharp contrast to the EP2s of $H_{2\times 2}$ displayed in Fig.~\ref{fig: EP2wPT 2x2}, where the pair annihilation is observed.

\section{
Details of the toy model
}
\label{sec: detail toy EP3PT}

Data displayed in Fig.~\ref{fig: EP3inPT} are obtained for the following toy model
\begin{align}
F(\omega,\bm{k}) = \omega\1 -H_0(\bm{k})-\Sigma(\omega)
\end{align}
with
\begin{align}
H_0(\bm{k}) &= 
\left(
\begin{array}{cccc}
0 & 1 & 0 & 0 \\
0 & 0 & 1 &  0 \\
0 & 0 & 0 & \delta_1 \\
3\sin k_2 & \cos k_1 +\cos k_2 -\frac{3}{2}& \delta_1& \delta_2 
\end{array}
\right),
\\
\Sigma(\omega) &= 0.1\omega\tanh(\omega)\mathrm{diag}(1,-1,1,-1).
\end{align}

For $\delta_1=0$, $H_0(\bm{k})$ is decoupled into $3\times3$ and $1\times 1$ blocks. The former reduces to a $3\times3$ Jordan block at $\bm{k}=\bm{0}$ and exhibits an EP$3$.

In order to compute the FM winding number $W_2$, we map the vector $\bm{d}$ to a Hermitian Hamiltonian $H_{\bm{d}}=d_1\sigma_1+d_2\sigma_2+d_3\sigma_3$ with Pauli matrices $\sigma_i$. 
By employing the method proposed in Ref.~\cite{Fukui_JPSJ2005}, we numerically obtain the Chern number of the Hermitian Hamiltonian $H_{\bm{d}}$ and thereby obtain $W_2$ (see also Appendix~\ref{sec: Hd}).

\section{
Doubling of nonlinear EP$n$s in terms of Hermitian Hamiltonians
}
\label{sec: Hd}

The FM winding numbers can be reinterpreted as topological invariants of Hermitian Hamiltonians, implying that the doubling of nonlinear EP$n$s can be understood in terms of the doubling of topological nodes of the corresponding Hermitian Hamiltonian.

With a given vector $\bm{d}$ that vanishes on an EP$n$ [e.g., see Eqs.~\eqref{eq: dvec nosymm} and \eqref{eq: dvec PT}], we consider a Hermitian Hamiltonian
\begin{align}
H_{\bm{d}}(\bm{q}) &=\sum_{i=1,2,\ldots,\mathrm{dim}\bm{d}} d_i\gamma_i
\end{align}
with Hermitian matrices $\gamma_i$ satisfying $\gamma_i \gamma_j +\gamma_j \gamma_i =2\delta_{ij}$ and $\bm{q}$ begin the vector in $\omega$-$\bm{k}$ space.
Here, the dimension of the $\omega$-$\bm{\bm{k}}$ space is equal to $\dim \bm{d}$ because EP$n$s emerge as zero-dimensional manifolds.
Sets of $\gamma$'s are obtained by taking the direct product of Pauli matrices $\sigma_i$ ($i=1,2,3$) and the identity matrix $\sigma_0$. For instance, a set of $\gamma_i$ is written as 
\begin{align}
\{\sigma_1,\sigma_2,\sigma_3 \},
\end{align}
for $2\times 2$ matrices and 
\begin{align}
\{\sigma_1\otimes\sigma_0,\sigma_2\otimes\sigma_0, \sigma_3\otimes\sigma_1,\sigma_3\otimes\sigma_2,\sigma_3\otimes\sigma_3  \},
\end{align}
for $4\times 4$ matrices. 
For $m_{\bm{d}}\times m_{\bm{d}}$ matrices, the number of possible $\gamma$ matrices is $2m_{\bm{d}}+1$.
The $\gamma$ matrices satisfy
\begin{align}
\label{eq: tr gamma1 ... gamma_2n+1}
 \mathrm{tr}\big(
 \gamma_1\gamma_2 \cdots \gamma_{2m_{\mathbb{d}}+1} 
 \big) &= (2\ii)^{m_{\mathbb{d}}}.
\end{align}

The vanishing of the vector $\bm{d}$ at the EP$n$ can be detected as the gap-closing of the Hermitian Hamiltonian $H_{\bm{d}}(\bm{q})$. This is because the eigenvalues of $H_{\bm{d}}(\bm{q})$ are given by $E_{\bm{d}}=\pm \sqrt{\lVert \bm{d} \rVert}$. When $\mathrm{dim}\bm{d}$ is odd, the symmetry class of $H_{\bm{d}}$ is class A because all possible $\gamma$'s are exhausted. 
In contrast, when $\mathrm{dim}\bm{d}$ is even, the symmetry class of $H_{\bm{d}}$ is class AIII due to the presence of one additional $\gamma$ matrix.

The above fact leads to the correspondence between the FM winding number and the Chern number (winding number) of the Hermitian Hamiltonian $H_{\bm{d}}$ when $\mathrm{dim}\bm{d}$ is odd (even).
Specifically, when $\mathrm{dim}\bm{d}$ is odd, the Chern number is written as
\begin{align}
\label{eq: C= int tr(QdQ)}
\mathcal{C}_{n_\mathrm{C}} &=-\frac{1}{2^{2n_\mathrm{C}+1}}\frac{1}{n_\mathrm{C}!} \Big(\frac{\ii }{2\pi }\Big)^{n_\mathrm{C}} \int \mathrm{tr} \Big[ Q (dQ)^{2n_\mathrm{C}}\Big]
\end{align}
with $n_\mathrm{C}=(\mathrm{dim}\bm{d}-1)/2$, $Q=\sum_i \hat{d}_i\gamma_i$ and $\hat{d}=\bm{d}/\lVert \bm{d} \rVert$.
Here, $Q$ is the flattened Hamiltonian associated with $H_{\bm{d}}$ whose eigenvalues are $\pm 1$ and $dQ$ denotes the exterior derivative of $Q$.
The integrant is rewritten as
\begin{widetext}

\begin{align}
\label{eq: QdQ Chern}
&\mathrm{tr} \Big[ Q (dQ)^{2n_\mathrm{C}}\Big]\nonumber \\
&=\mathrm{sgn}
\left(
\begin{array}{ccc}
1 &\cdots & 2n_\mathrm{C} \\
\mu_1 & \cdots & \mu_{2n_\mathrm{C}} 
\end{array}
\right)
\mathrm{tr}
[\gamma_{i_1}\cdots\gamma_{i_{2n_\mathrm{C}+1}}]
\big(
\hat{d}_{i_1}\partial_{\mu_1}
\hat{d}_{i_2}
\cdots
\partial_{\mu_{2n_\mathrm{C}}}\hat{d}_{i_{2n_\mathrm{C}+1}}
\big)
d^{2n_{\mathrm{C}}}\bm{q}
\nonumber \\
&=
\mathrm{sgn}
\left(
\begin{array}{ccc}
1 &\cdots & 2n_\mathrm{C} \\
\mu_1 & \cdots & \mu_{2n_\mathrm{C}} 
\end{array}
\right)
\mathrm{sgn}
\left(
\begin{array}{ccc}
1 &\cdots & 2n_\mathrm{C}+1 \\
i_1 & \cdots & i_{2n_\mathrm{C}+1} 
\end{array}
\right)
\mathrm{tr}
[\gamma_{1}\cdots\gamma_{2n_\mathrm{C}+1}]
\big(
\hat{d}_{i_1}\partial_{\mu_1}
\hat{d}_{i_2}
\cdots
\partial_{\mu_{2n_\mathrm{C}}}\hat{d}_{i_{2n_\mathrm{C}+1}}
\big)
d^{2n_{\mathrm{C}}}\bm{q}
\nonumber \\
&=
(2\ii)^{n_{\mathrm{C}}}
\mathrm{sgn}
\left(
\begin{array}{ccc}
i_2 &\cdots & i_{2n_\mathrm{C}+1} \\
\sigma(i_2) & \cdots & \sigma(i_{2n_\mathrm{C}+1}) 
\end{array}
\right)
\mathrm{sgn}
\left(
\begin{array}{ccc}
1 &\cdots & 2n_\mathrm{C}+1 \\
i_1 & \cdots & i_{2n_\mathrm{C}+1} 
\end{array}
\right)
\big(
\hat{d}_{i_1}\partial_{1}
\hat{d}_{\sigma(i_2)}
\cdots
\partial_{2n_\mathrm{C}}\hat{d}_{\sigma(i_{2n_\mathrm{C}+1})}
\big)
d^{2n_{\mathrm{C}}}\bm{q}
\nonumber \\
&=
(2\ii)^{n_{\mathrm{C}}}
\mathrm{sgn}
\left(
\begin{array}{cccc}
1   &2             &\cdots & 2n_\mathrm{C}+1 \\
i_1 &\sigma(i_2)  &\cdots & \sigma(i_{2n_\mathrm{C}+1}) 
\end{array}
\right)
\big(
\hat{d}_{i_1}\partial_{1}
\hat{d}_{\sigma(i_2)}
\cdots
\partial_{2n_\mathrm{C}}\hat{d}_{\sigma(i_{2n_\mathrm{C}+1})}
\big)
d^{2n_{\mathrm{C}}}\bm{q},
\end{align}
where the summation is taken over repeated indices. 
The symbol ``$\mathrm{sgn}(\cdots)$" denotes $1$ ($-1$) when the second row is an even (odd) permutation of the first row. 
The symbol ``$d^{2n_{\mathrm{C}}}\bm{q}$" is defined as $d^{2n_{\mathrm{C}}}\bm{q}=dq_1dq_2\ldots dq_{2n_{\mathrm{C}}}$. 
Here $\sigma(i)$ in the fourth line denotes the permutation of indices so that  
\begin{align*}
\mathrm{sgn}
\left(
\begin{array}{ccc}
1 &\cdots & 2n_\mathrm{C} \\
\mu_1 & \cdots & \mu_{2n_\mathrm{C}} 
\end{array}
\right)
\big(
\hat{d}_{i_1}\partial_{\mu_1}
\hat{d}_{i_2}
\cdots
\partial_{\mu_{2n_\mathrm{C}}}\hat{d}_{i_{2n_\mathrm{C}+1}}
\big) 
&=
\sum_\sigma
\mathrm{sgn}
\left(
\begin{array}{ccc}
i_2 &\cdots & i_{2n_\mathrm{C}+1} \\
\sigma(i_2) & \cdots & \sigma(i_{2n_\mathrm{C}+1}) 
\end{array}
\right)
\big(
\hat{d}_{i_1}\partial_{1}
\hat{d}_{\sigma(i_2)}
\cdots
\partial_{2n_\mathrm{C}}\hat{d}_{\sigma(i_{2n_\mathrm{C}+1})}
\big)
\end{align*}
holds. The symbol ``$\sum_\sigma$" denotes the summation of all possible permutations.
\end{widetext}
From the third to the fourth line of Eq.~\eqref{eq: QdQ Chern}, we have used Eq.~\eqref{eq: tr gamma1 ... gamma_2n+1}. 
Substituting Eq.~\eqref{eq: QdQ Chern} into Eq.~\eqref{eq: C= int tr(QdQ)}, we obtain
\begin{align}
\label{eq: (-1)^n C=...}
&(-1)^{n_{\mathrm{C}}+1}\mathcal{C}_{n_\mathrm{C}} \\
&=\frac{(2n_{\mathrm{C}})!}{\pi^{n_{\mathrm{C}}}  n_{\mathrm{C}}! 2^{2n_{\mathrm{C}}+1}}
 \int d^{2n_{\mathrm{C}}}\bm{q} \epsilon^{i_1\ldots i_{2n_{\mathrm{C}}}}
f_{i_1\ldots i_{2n_{\mathrm{C}}}}\nonumber \\
&=W_{2n_{\mathrm{C}}}.
\end{align}
From the second to the third lines, we have used Eqs.~\eqref{eq: W no symm} and \eqref{eq: A_(2M)}.
This relation indicates the correspondence between FM winding numbers $W_{2n_{\mathrm{C}}}$ and Chern numbers $\mathcal{C}_{n_\mathrm{C}}$ of Hermitian Hamiltonians.

When $\mathrm{dim}\bm{d}$ is even, the winding number of $H_{\bm{d}}$ is written as
\begin{align}
\label{eq: W= int tr(gamma QdQ)}
&\mathcal{W}_{2n_\mathrm{W}+1} \nonumber \\ &=\frac{(-1)^{n_\mathrm{W}} n_\mathrm{W}! }{2 (2\pi \ii)^{n_\mathrm{W}+1} (2n_\mathrm{W}+1)!}
\int \mathrm{tr} \Big[\gamma_{2n_\mathrm{W}+3} Q (dQ)^{2n_\mathrm{W}+1}\Big]
\end{align}
with $n_\mathrm{W}=\mathrm{dim}\bm{d}/2-1$, $Q=\sum_{i=1,\ldots,2n_\mathrm{W}+2} \hat{d}_i\gamma_i$ and $\hat{d}=\bm{d}/\lVert \bm{d} \rVert$.

The integrant is rewritten as
\begin{widetext}
\begin{align}
\label{eq: QdQ Wind}
&\mathrm{tr} \Big[\gamma_{2n_\mathrm{W}+3} Q (dQ)^{2n_\mathrm{W}+1}\Big]
\nonumber \\
&=\mathrm{sgn}
\left(
\begin{array}{ccc}
1 &\cdots & 2n_\mathrm{W}+1 \\
\mu_1 & \cdots & \mu_{2n_\mathrm{W}+1} 
\end{array}
\right)
\mathrm{tr}
[\gamma_{2n_\mathrm{W}+3} \gamma_{i_1}\cdots\gamma_{i_{2n_\mathrm{W}+2}}]
\big(
\hat{d}_{i_1}\partial_{\mu_1}
\hat{d}_{i_2}
\cdots
\partial_{\mu_{2n_\mathrm{W}+1}}\hat{d}_{i_{2n_\mathrm{W}+2}}
\big)
d^{2n_{\mathrm{W}}+1}\bm{q}
\nonumber \\
&=
\mathrm{sgn}
\left(
\begin{array}{ccc}
1 &\cdots & 2n_\mathrm{W}+1 \\
\mu_1 & \cdots & \mu_{2n_\mathrm{W}+1} 
\end{array}
\right)
\mathrm{sgn}
\left(
\begin{array}{ccc}
1 &\cdots & 2n_\mathrm{W}+2 \\
i_1 & \cdots & i_{2n_\mathrm{W}+2} 
\end{array}
\right)
\mathrm{tr}
[\gamma_{1}\cdots\gamma_{2n_\mathrm{W}+3}]
\big(
\hat{d}_{i_1}\partial_{\mu_1}
\hat{d}_{i_2}
\cdots
\partial_{\mu_{2n_\mathrm{W}+1}}\hat{d}_{i_{2n_\mathrm{W}+2}}
\big)
d^{2n_{\mathrm{W}}+1}\bm{q}
\nonumber \\
&=
(2\ii)^{n_{\mathrm{W}}+1}
\mathrm{sgn}
\left(
\begin{array}{ccc}
i_2 &\cdots & i_{2n_\mathrm{W}+2} \\
\sigma(i_2) &\cdots & \sigma(i_{2n_\mathrm{W}+2}) 
\end{array}
\right)
\mathrm{sgn}
\left(
\begin{array}{ccc}
1 &\cdots & 2n_\mathrm{W}+2 \\
i_1 & \cdots & i_{2n_\mathrm{W}+2} 
\end{array}
\right)
\big(
\hat{d}_{i_1}\partial_{1}
\hat{d}_{\sigma(i_2)}
\cdots
\partial_{2n_\mathrm{W}+1}\hat{d}_{\sigma(i_{2n_\mathrm{W}+2})}
\big)
d^{2n_{\mathrm{W}}+1}\bm{q} 
\nonumber  \\
&=
(2\ii)^{n_{\mathrm{W}}+1}(2n_\mathrm{W}+1)!
\mathrm{sgn}
\left(
\begin{array}{cccc}
1   & 2   &\cdots & 2n_\mathrm{W}+2 \\
i_1 & i_2 &\cdots & i_{2n_\mathrm{W}+2} 
\end{array}
\right)
\big(
\hat{d}_{i_1}\partial_{1}
\hat{d}_{i_2}
\cdots
\partial_{2n_\mathrm{W}+1}\hat{d}_{i_{2n_\mathrm{W}+2}}
\big)
d^{2n_{\mathrm{W}}+1}\bm{q},
\end{align}
in a similar way as Eq.~\eqref{eq: QdQ Chern}.
%
\end{widetext}
%
Substituting Eq.~\eqref{eq: QdQ Wind} into Eq.~\eqref{eq: W= int tr(gamma QdQ)}, we have
\begin{align}
\label{eq: (-1)^n W=...}
&(-1)^{n_{\mathrm{W}}}\mathcal{W}_{2n_\mathrm{W}+1} \nonumber \\
&=
\frac{n_\mathrm{W}! }{2 \pi^{n_\mathrm{W}+1} }
 \int d^{2n_{\mathrm{W}}+1}\bm{q} \epsilon^{i_1\ldots i_{2n_{\mathrm{W}}+1}}
f_{i_1\ldots i_{2n_{\mathrm{W}}+1}}
\nonumber \\
&=W_{2n_{\mathrm{W}}+1}.
\end{align}
From the second to the third lines, we have used Eqs.~\eqref{eq: W no symm} and \eqref{eq: A_(2M+1)}.
This relation indicates the correspondence between FM winding numbers $W_{2n_{\mathrm{W}}+1}$ and winding numbers $\mathcal{W}_{2n_\mathrm{W}+1}$ of Hermitian Hamiltonians.

\end{document}